# Rapid and cost-effective evaluation of bacterial viability using fluorescence spectroscopy

Authors: Fang Ou[1,2,*], Cushla McGoverin[1,2], Simon Swift[3], Frédérique Vanholsbeeck[1,2]

[1] Department of Physics, The University of Auckland, Auckland, New Zealand

[2] The Dodd-Walls Centre for Photonic and Quantum Technologies, New Zealand

[3] School of Medical Sciences, The University of Auckland, Auckland, New Zealand

**\*Corresponding author**

Email: fou521@aucklanduni.ac.nz

## Abstract
The fluorescence spectra of bacterial samples stained with SYTO 9 and propidium iodide (PI) were used to monitor bacterial viability. Stained mixtures of live and dead *Escherichia coli* with proportions of live:dead cells varying from 0 to 100% were measured using the optrode, a cost effective and convenient fibre-based spectroscopic device. We demonstrated several approaches to obtaining the proportions of live:dead *E. coli* in a mixture of both live and dead, from analyses of the fluorescence spectra collected by the optrode. To find a suitable technique for predicting the percentage of live bacteria in a sample, four analysis methods were assessed and compared: SYTO 9:PI fluorescence intensity ratio, an adjusted fluorescence intensity ratio, single-spectrum support vector regression (SVR) and multi-spectra SVR. Of the four analysis methods, multi-spectra SVR obtained the most reliable results and was able to predict the percentage of live bacteria in $10^8$ bacteria/mL samples between *c.* 7% and 100% live, and in $10^7$ bacteria/mL samples between *c.* 7% and 73% live. By demonstrating the use of multi-spectra SVR and the optrode to monitor *E. coli* viability, we raise points of consideration for spectroscopic analysis of SYTO 9 and PI and aim to lay the foundation for future work that use similar methods for different bacterial species.

## Introduction

Monitoring bacterial viability is an important task in many fields of microbiological study including the monitoring of food safety and public health. Current standard assessments rely largely on the agar plate count method. A count of the viable bacteria in the sample is obtained via enumeration of the colony forming units (CFU) following an incubation period, with the assumption that each CFU grew from one bacterium of the sample [1]. Due to the need for incubation, the agar plate count process requires 1 to 5 days [1, 2]. In addition, only the cells that can form colonies under the conditions of the experiment will be counted, thus providing no indication of dead bacteria or viable but non-culturable (VBNC) cells [1, 3, 4].

A common alternative method of analysis involves using fluorescent dyes SYTO 9 and propidium iodide (PI) which differentially stain live and dead bacteria. The signals from the dyes are typically measured using fluorescence microscopy, fluorescence-based microplate readers or flow cytometry. Fluorescence microscopy provides direct morphological information of individual cells, but it has a small field of view thus the analysis of large sample volumes become time consuming [5, 6]. Compared to microscopy, fluorescence-based microplate readers and flow cytometry (FCM) have superior ease of use, as many of its operations can be automated and performed in parallel [7, 8]. However, the accuracy of microplate reader and FCM measurements depend on the sensitivity of the instrument and the quality of the optics, which increases with cost [9–11]. In addition, the equipment is bulky, and often requires operation by trained technicians.

In pursuit of a rapid, convenient and cost-effective method for monitoring bacterial viability, the optrode has proven promising [12]. The optrode is a fibre-based spectroscopic system that can accurately quantify the fluorescence signals from a sample. The quantitative fluorescence spectra obtained by the optrode are corrected for variations in excitation illumination and signal integration times, thus enabling the comparison of measurements from samples across a wide concentration range. The optrode measures fluorescence across the entire visible range and the sample exposure time can be varied between 8 ms to 10 s. This allows detailed characterisation of various fluorophores and how their signal changes in different experimental environments.

In this study the optrode was used to accurately measure fluorescence emissions from bacterial samples stained with both SYTO 9 and PI. Initially to calibrate the spectral profile to percentage of live bacteria present, the ratio of the SYTO 9:PI peak intensity described in the LIVE/DEAD BacLight Bacterial Viability Kits manual was used [13]. However, our results showed that the linear relationship between the SYTO 9:PI intensity ratio and the percentage of live bacteria is not reliable, and becomes variable particularly above $c.$ 60% live. This observation agrees with previous studies where the SYTO 9:PI dye ratios were obtained from fluorescence-based microplate reader measurements [14–16]. We show alternative approaches to the analysis of quantitative fluorescence spectral data, to obtain with higher accuracy a measure of the percentage of live *Escherichia coli* present in a mixture of both live and dead cells.

## Materials and methods
### Bacterial culture conditions
As a model, *Escherichia coli* strain ATCC 25922 (American Type Culture Collection, Virginia, USA) was used in all experiments. *E. coli* was grown overnight then sub-cultured at 20× dilution and incubated for 1 h to reach an optical density (OD) between 0.5 and 0.6 at 600 nm (path length 1 cm), representing $c.$ $4 \times 10^8$ CFU/mL. All broth cultures were incubated at 37°C in Difco tryptic soy broth (TSB; Fort Richard Laboratories, Auckland, New Zealand) and aerated with orbital shaking at 200 rpm.

### Live:dead bacterial mixtures
Live and dead bacterial mixtures were obtained using previously described methods [17]. Briefly, the sub-cultured *E. coli* in the exponential growth phase were harvested via centrifugation (4302 × $g$, 10 min, 21°C), followed by removal of supernatant and resuspension of the pellet in 3 mL of saline. Aliquots of the resuspended cells were diluted 1:9 mL in either saline for live cells, or 70% isopropanol for dead cells. Each bacterial suspension was shaken at 200 rpm at 28°C for 1 h. The live and dead bacterial cells were harvested via centrifugation (4302× $g$, 10 min, 21°C), followed by removal of the supernatant and resuspension of the pelleted cells in 20 mL of saline. After three washing cycles, each of the live and dead cell suspensions was diluted to achieve a concentration of $c.$ $1 \times 10^8$ bacteria/mL; equivalent to diluting the sample to a final OD of 0.168±0.063 at 600 nm. These live and dead bacterial suspensions were used either directly or diluted to $1 \times 10^7$ bacteria/mL, to prepare mixtures with live:dead proportions corresponding to 0, 2.5, 5, 10, 25, 50, 75, 100 % live bacteria.

### Fluorescent dye staining and treatment of unbound dyes
BacLight LIVE/DEAD Bacterial Viability and Counting Kits (Invitrogen, Molecular Probes, Carlsbad, CA, USA; L34856) were used in all experiments. The kit comprises a vial of microsphere suspension, and two nucleic acid dyes SYTO 9 and propidium iodide (PI) that label live and dead bacteria, respectively. For each experiment, saline was used as diluent to make working solutions of SYTO 9 and PI with concentrations of 33.4 μM and 0.4 mM, respectively. The microsphere suspension was sonicated (SC-120 sonicator, Sonicor, NY, USA) for 10 min in a water bath and gently vortexed at 500 rpm prior to use. For each sample, 50 μL each of the SYTO 9 and PI working solutions, and 10 μL of the microsphere suspension, were aliquoted into an empty microcentrifuge tube. Subsequently, 900 µL of each bacterial sample was added to each tube then gently vortexed at 500 rpm in the dark for 15 min at room temperature, to allow dye-bacteria binding.

### Flow cytometry protocols

Reference measurements of all samples were obtained using a LSR II Flow Cytometer (BD Biosciences, San Jose, CA, USA), using previously described methods [17]. In brief, samples were excited with 488 nm laser with 20 mW power. SYTO 9 fluorescence was collected using a 505 nm longpass filter and a 530/30 nm bandpass filter. PI fluorescence was collected using a 685 nm longpass filter and a 695/40 nm bandpass filter. The threshold was set to side scatter at 200 and the bead count per second was plotted to monitor for any disturbance or blockages. The measurement duration was 150 s and the flow rate was set to *c.* 6 µL/min. Gating was done in the red fluorescence vs green fluorescence dot plot to obtain the percentage of live and dead bacteria in each sample.

### Optrode protocols

Fluorescence dye emission spectra were recorded from bacterial samples using a fibre-based spectroscopic system (optrode) [12]. Excitation is achieved by a 473 nm solid state laser with *c.* 10 mW power. Signal variation caused by photobleaching is minimised by synchronisation of the laser shutter with the spectrometer using a data acquisition (DAQ) card. Using a 2x2 fibre coupler, the laser irradiates both the sample and a photodiode that monitors the laser power fluctuations. Fluorescence excitation and detection are achieved by a single fibre probe (200 um diameter, 0.22 NA; Thorlabs Inc., Newton, NJ, USA). A 495 nm longpass filter removes the excitation line before signals reach the Ocean Optics QE65000 CCD spectrometer.

### Spectra acquisition and pre-processing

*N*=56 standard bacterial samples (*n*=159 optrode measurements) and *N*=27 test set samples (*n*=80 optrode measurements) were analysed, on average each sample was measured three times. The duration of each optrode measurement was 10 s and each consisted of 500 spectra collected consecutively using a 20 ms integration time per spectrum; these optrode measurements allowed photobleaching behaviour to be observed. The instrument dark noise was removed, and all spectra were normalised to 8 ms integration time and 10 mW laser power. The averaged background spectrum obtained from saline was subtracted from each sample spectrum. Occasionally anomalous optrode measurements, with noticeably higher or lower intensity compared to the rest of the optrode measurements collected from the sample, were obtained. The occurrences of anomalous optrode measurements were substantially reduced by thorough cleaning of the probe. The anomalous optrode measurements were excluded from analysis, noted, and are not included in the optrode measurement count (*n*) stated in the beginning of this section. The remaining fluorescence spectra were subsequently mean-centred with respect to the average of the appropriate training set spectrum, and from each optrode measurement, up to 5 spectra were chosen from the series of 500 and used in algorithms to predict the percentage of live and dead bacteria measured by FCM.

### Fluorescence photobleaching rates

Fluorescence photobleaching rates were obtained from dual-stained bacterial samples at concentration of *c.* $10^8$ bacteria/mL. In the dual stained dead bacterial samples, while the PI will be mostly bound (mB) and SYTO 9 will be mostly unbound (mU), it is likely that some PI will remain unbound and some SYTO 9 will be bound. On the other hand, it is expected that in the dual stained live bacterial samples, some SYTO 9 will remain unbound. The mB-SYTO 9 and mU-PI photobleaching half-lives were obtained from fluorescence emission of dual-stained live bacteria. Vice versa, the photobleaching half-lives of mU-SYTO 9 and mB-PI were obtained from fluorescence emission of dual-stained dead bacteria. The fluorescence intensity of the dyes over a 10 s duration were measured using the optrode and modelled by a double exponential to find the photobleaching half-lives of the dyes in each situation (*N*=6 bacterial samples, *n*=17 optrode measurements, on average three per sample). The t-test was performed in Python to evaluate whether the photobleaching half-life of the dyes in bound or unbound states were significantly different.

### Data analysis

To find a suitable technique for predicting the percentage of live bacteria in a sample, four analysis methods were assessed and compared: SYTO 9:PI intensity ratio, adjusted dye ratio [18], single-spectrum support vector regression (SVR), and multi-spectra SVR. The four methods were evaluated by external validation using test set samples. The $R^2$, standard error, root mean square error (RMSE) and explained variance were found.

### Dye ratio

In this work, the 'dye ratio' refers to the SYTO 9:PI integrated intensity ratio described in the LIVE/DEAD BacLight Bacterial Viability Kits manual, shown on the left-hand side of Equation (1):

$$\frac{SYTO\ 9}{PI} \propto \%live \qquad (1)$$

Where *SYTO 9* and *PI* represents the integrated intensity of SYTO 9 and PI, respectively. *%live* is the percent of live bacteria in the sample, measured by flow cytometry. The regions of intensity integration corresponded to the fluorescence peak of the dyes, which were between 509 - 529 nm for SYTO 9 and 609 - 629 nm for PI. This dye ratio was calculated then compared to the percent of live bacteria in a sample.

### Adjusted dye ratio

The adjusted dye ratio is derived assuming the ideal behaviour of the dyes, wherein SYTO 9 and PI stains the live and dead cells, respectively. The relationship can be expressed as below:

$$\frac{\%live}{100 - \%live} \propto \frac{SYTO\ 9}{PI} \qquad (2)$$

The above expression (1) can be rearranged to:

$$\%live \propto \frac{100 \times SYTO\ 9/PI}{1 + SYTO\ 9/PI} \qquad (3)$$

The right-hand side of Equation (2) is referred to as the *adjusted dye ratio*, and this value is calculated and compared to the percentage of live cells present in each sample.

### Single-spectrum SVR

The first spectrum from each optrode measurement was used as input for the SVR algorithm, which maps the input data onto a high-dimensional feature space, then constructs a regression model in the feature space [19]. Unlike least squares regression which aims to only minimise the observed training error, SVR seeks to minimise both the observed training error and a regularisation value that controls the complexity of the model [20]. The ε-SVR from Python's Scikit-learn library [21] was applied using the linear kernel. The generalisation ability of ε-SVR is controlled by two parameters ε and C, which defines the margin of tolerance and the penalty factor, respectively.

Grid search and group K-fold cross validation (GKCV) were implemented to find the best estimator values for the ε and C parameters that minimised the mean squared error of the predictions. Grid search was applied to search over various parameter values of both ε and C to find the estimators that minimised the mean squared error of the predictions. This process was optimised by GKCV, where the spectral training dataset was split into groups according to the *M* experiments performed to collect the data at each bacterial concentration, which were 3 and 4 for the $10^7$ and $10^8$ bacteria/mL samples, respectively. Then, *M* iterations of GKCV were performed and in each iteration. One group was held out as the internal test set while the remaining were used as the training set. The final combination of ε and C values that minimised the RMSE of the predictions were chosen.

### Multi-spectra SVR

Various combinations of 2, 3 or 5 spectra obtained at different time points in one optrode measurement were concatenated and analysed by SVR. Five timepoints were investigated that corresponded to: the first 20 ms following fluorescence excitation, and the photobleaching half-lives of mB-SYTO 9, mB-PI, mU-SYTO 9 and mU-PI. To find the spectral combination which returned the lowest RMSE, initial assessments by GKCV were completed using default ε and C values (of 0.1 and 1, respectively). Subsequently, the ε and C parameters of the SVR model that used the chosen spectral combination were optimised, using grid search and GKCV.

## Results

### Fluorescence spectra and the SYTO 9 and PI peaks

The SYTO 9 and PI emissions of bacterial samples vary as the concentration of live and dead bacteria in the sample changes. Figure 1 shows spectra obtained from *E. coli* samples at concentration of $10^8$ bacteria/mL, stained using either SYTO 9 alone, PI alone, or both SYTO 9 and PI. The samples contain varying proportions of live and dead cells, and the fluorescence peaks of SYTO 9 and PI are observed near 509 – 529 nm and 609 – 629 nm, respectively. Comparing the spectra of samples stained using SYTO 9 only: the SYTO 9 peak intensity more than doubles from the 100% live sample to the 100% dead sample. On the other hand, there is a slight peak intensity decrease of *c.* 13% from the 100% live sample to the 50% live sample. Samples stained using only PI showed lower peak intensity than those stained using only SYTO 9, however the PI fluorescence is enhanced in the presence of SYTO 9.

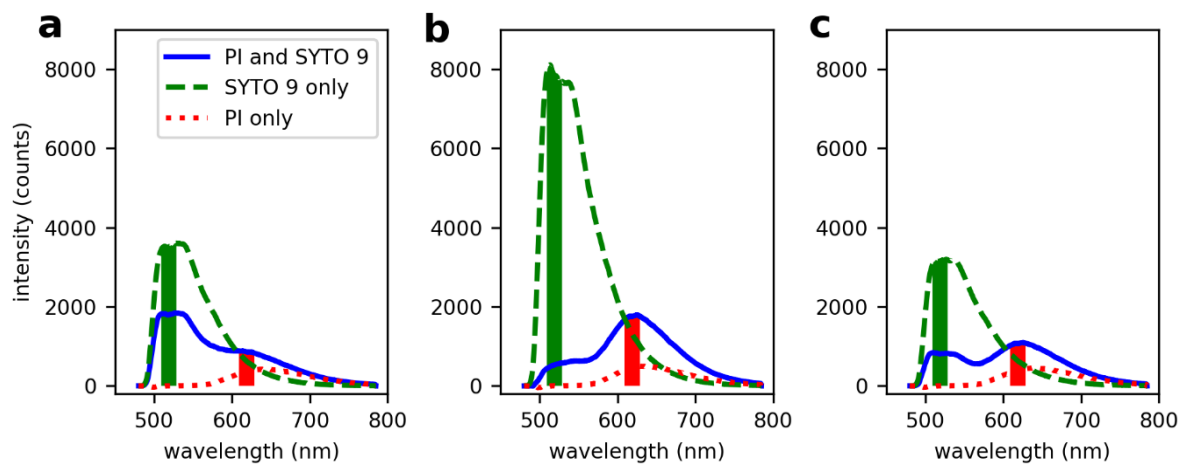

**Fig. 1** Exemplar spectra showing the difference in spectral profile obtained from samples containing $10^8$ bacteria/mL of live (a), dead (b) and 50:50 live and dead (c) bacteria. The bacterial samples were stained using PI only, SYTO 9 only or both PI and SYTO 9. The shaded regions near 510 nm and 610 nm represents the intervals used to obtain the integrated intensity of SYTO 9 and PI, respectively.

### Fluorescence photobleaching rates

To obtain the photobleaching half-lives of SYTO 9 and PI, dual stained live or dead bacterial suspensions at concentration of $10^8$ bacteria/mL were measured using the optrode. The mean photobleaching half-lives of SYTO 9 compared to PI, and in their mostly bound (mB) versus mostly unbound (mU) states were not significantly different (p-values > 0.05). Nonetheless, the photobleaching half-lives provide an indication of time points to inspect for spectral changes that occur over time, which is useful in the multi-spectra SVR analysis. The average photobleaching half-life of mB- or mU- SYTO 9 and PI are summarised in Table 1.

|  | Photobleaching half-life (ms) | |
| --- | --- | --- |
| State of dye | SYTO 9 | PI |
| Mostly bound (mB) | 1244 ± 97 | 1124 ± 115 |
| Mostly unbound (mU) | 1472 ± 165 | 1084 ± 29 |

**Table 1** The average photobleaching half-lives of SYTO 9 and PI ± the standard error, obtained from dual stained live or dead bacterial samples with concentration of $10^8$ bacteria/mL. The photobleaching half-lives for mostly bound-SYTO 9 and mostly unbound-PI were obtained from dual stained live bacteria, whereas those of mostly unbound-SYTO 9 and mostly bound-PI were obtained from dual stained dead bacteria.

### Regression models: dye ratio, adjusted dye ratio, single-spectrum SVR and multi-spectra SVR

The spectral training data for $10^8$ bacteria/mL samples (*N*=32 samples; *n*=91 optrode measurements, on average three per sample) were obtained from four experiments. The spectral training data for $10^7$ bacteria/mL samples (*N*=24 samples; *n*=68 measurements, on average three per sample) were obtained from three experiments. To establish a method for correlating the spectral changes to variations in proportions of live:dead bacteria, four regression models were compared which used: the dye ratio, adjusted dye ratio, single-spectrum SVR and multi-spectra SVR.

Figure 2 shows the dye ratio and adjusted dye ratio as a function of the percentage of live bacteria in $10^8$ bacteria/mL training samples. Overall, at $10^8$ bacteria/mL, the adjusted dye ratio has a much more linear relationship with percentage of live bacteria, compared to the dye ratio. The relationship between the dye ratio and percentage of live bacteria appears nonlinear beyond *c.* 60% live. On the other hand, the adjusted dye ratio maintains a roughly linear relationship with the percentage of live bacteria, but the variations in results are greater beyond 60%. The $R^2$ were -1.1 and 0.67 for the dye ratio and adjusted dye ratio models, respectively, with the negative $R^2$ indicating that the fit is worse than the mean of the data [21]. Analysis using the dye ratio and adjusted dye ratio were not useful for $10^7$ bacteria/mL samples as the results were highly variable.

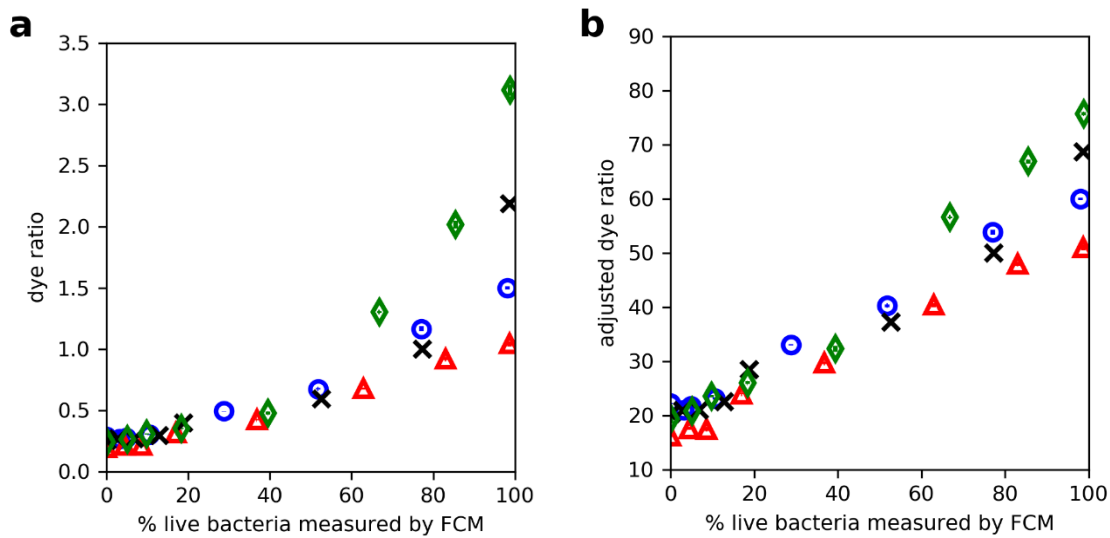

**Fig. 2** The (a) dye ratio and (b) adjusted dye ratio as a function of the percentage of live bacteria in the $10^8$ bacteria/mL training samples. Four types of markers represent the separate experiments from which the data were collected.

Fluorescence spectra obtained at various time points corresponding to the photobleaching half-lives of the dyes in mB or mU states were concatenated with the spectrum recorded in the first 20 ms ($S_1$) and analysed in multi-spectra SVR (Table 2). For the analysis of $10^8$ bacteria/mL samples, a combination of $S_1$ and the spectrum taken at the photobleaching half-live of mB-SYTO 9 (1244 ms) returned the lowest error value in GKCV. Single spectrum SVR analysis of $10^7$ bacteria/mL samples returned the lowest RMSE for $10^7$ bacteria/mL samples, slightly lower than using a combination of $S_1$ and the spectrum taken at the photobleaching half-life of mB-PI (1124 ms), which returned the lowest error value of multi-spectra SVR.

| Number of spectra used | Time points used | Pre-optimised RMSE results for $10^8$ bacteria/mL samples | Pre-optimised RMSE results for $10^7$ bacteria/mL samples |
|---|---|---|---|
| 1 | $S_1$ (single-spectrum SVR) | 20.0 | 21.6 |
| 2 | $S_1$, mB-SYTO 9 $T_{1/2}$ | 19.85 | 23.55 |
|   | $S_1$, mB-PI $T_{1/2}$ | 20.73 | 22.87 |
|   | $S_1$, mU-SYTO 9 $T_{1/2}$ | 20.67 | 23.29 |
|   | $S_1$, mU-PI $T_{1/2}$ | 20.61 | 22.92 |
| 3 | $S_1$, mB-SYTO 9 $T_{1/2}$, mB-PI $T_{1/2}$ | 20.70 | 23.32 |
|   | $S_1$, mU-SYTO 9 $T_{1/2}$, mU-PI $T_{1/2}$ | 22.07 | 22.96 |
| 5 | $S_1$, mB-SYTO 9 $T_{1/2}$, mB-PI $T_{1/2}$, mU-SYTO 9 $T_{1/2}$, mU-PI $T_{1/2}$ | 21.94 | 23.29 |

**Table 2** Combinations of spectra used in multi-spectra SVR analysis. $S_1$ represents the spectrum recorded in the first 20 ms of the measurement, and $T_{1/2}$ represents the measured photobleaching half-life of the associated fluorophore in its mostly bound (mB) or mostly unbound (mU) state. The RMSE values of $10^8$ and $10^7$ bacteria/mL samples were obtained via GKCV prior to parameter optimisation of the SVR algorithm.

The single-spectrum SVR and multi-spectra SVR regression models both obtained linear relationships for modelling the percentage of live bacteria, when using $10^8$ or $10^7$ bacteria/mL training samples. Grid search and GKCV returned 0.5 and 0.001 as the optimal values for SVR hyper parameters C and ε, respectively. The RMSE and explained variance of the predictions are summarised in Table 3.

| Property modelled | Regression models | $10^8$ bacteria/mL samples | | $10^7$ bacteria/mL samples | |
|---|---|---|---|---|---|
|  |  | RMSE | Explained variance | RMSE | Explained variance |
| % live | Single-spectrum SVR | 19.95 | 0.85 | 21.56 | 0.91 |
|  | Multi-spectra SVR | 19.79 | 0.87 | 22.86 | 0.91 |

**Table 3** The RMSE and explained variance of the single-spectrum SVR and multi-spectra SVR models in modelling the percentage of live bacteria in training samples. Multi-spectra SVR model for the analysis of $10^8$ bacteria/mL samples used a combination of the spectrum recorded in the first 20 ms ($S_1$) and the spectrum taken at the photobleaching half-live of mostly bound-SYTO 9 (1244 ms); multi-spectra SVR model for the analysis of $10^7$ bacteria/mL samples used a combination of the spectrum taken at $S_1$ and the spectrum taken at the photobleaching half-life of mostly bound-PI (1124 ms).

### Validation of SVR models using test set samples

The single-spectrum and multi-spectra SVR models were validated and compared using external test set samples ($N$=27 samples; $n$=80 optrode measurements, on average three per sample) collected from two blind experiments. The predictions obtained from multiple spectral measurements of each sample were averaged then compared to the percentage of live bacteria measured by FCM. Both the single-spectrum and multi-spectra SVR models predicted the percentage of live bacteria in the samples reasonably well. Results of the comparison between the two SVR models are shown in Table 4 and overall, the predictions obtained using multi-spectra SVR deviated less from the expected values compared to those obtained using single-spectrum

SVR. Both SVR models performed better in predicting the percentage of live bacteria in samples with concentration of $10^8$ than $10^7$ bacteria/mL. It was observed that the percentage live predictions for $10^7$ bacteria/mL samples containing less than *c.* 7% or more than 73% returned percentage values that were negative or above 100%. These were considered as invalid predictions and excluded.

| SVR model | Time points used in SVR model | Concentration of samples (bacteria/mL) | Standard error | % of samples with predictions within 2 SE | No. of samples with invalid predictions |
|---|---|---|---|---|---|
| Single-spectrum | $S_1$ | $10^8$ | 5.5 | 100 | 0 |
| | $S_1$ | $10^7$ | 16.1 | 100 | 3 |
| Multi-spectra | $S_1$, mB-SYTO 9 $T_{1/2}$ | $10^8$ | 4.7 | 91 | 0 |
| | $S_1$, mB-PI $T_{1/2}$ | $10^7$ | 13.1 | 100 | 3 |

**Table 4** Evaluation of the single-spectrum and multi-spectra SVR models in predicting the percentage of live bacteria in $10^8$ and $10^7$ bacteria/mL samples. Invalid predictions refer to model outputs that were negative or above 100%.

Predictions of the percentage of live bacteria in test set samples using multi-spectra SVR are shown in Figure 3. All predictions lie within the 95% confidence interval of each model, or $\pm 2$ standard errors of the 1:1 line [22], except for one prediction of a $10^8$ bacteria/mL sample containing low percentage of live cells that was below 7%.

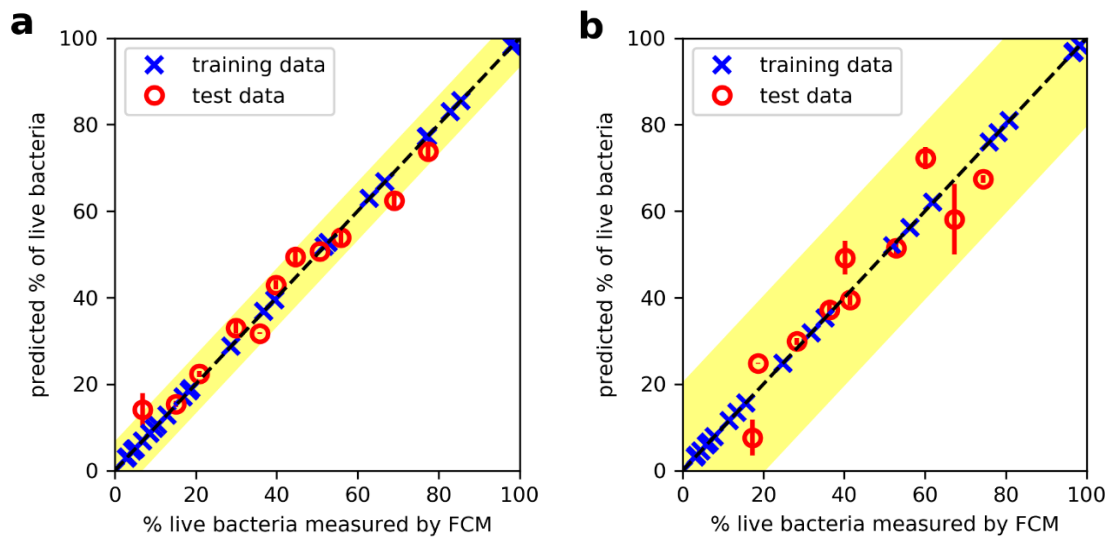

**Fig. 3** The percentage of live bacteria predicted using the multi-spectra SVR model compared to that measured by FCM. The samples contain a mixture of live and dead *E. coli* cells, with concentration of (a) $10^8$ or (b) $10^7$ bacteria/mL. The dashed line marks the 1:1 relationship between the predicted percent live and that measured using FCM. The shaded area represents the region of plus or minus two standard errors of the 1:1 line. The standard error in replicate measurements are represented by the vertical and horizontal error bars. A few $10^7$ bacteria/mL samples returned invalid predictions that were negative or above 100%, which were excluded.

## Discussion

In this study, we showed several approaches to obtaining the ratio of live:dead *E. coli* in a mixture of both live and dead, from analyses of its fluorescence spectra. Spectral measurements were obtained using a portable and cost-effective lab-built optrode system that accurately quantifies fluorescence signals in near real time. Fluorescence was measured from SYTO 9- and PI- stained *E. coli* samples with varying proportions of live and dead cells, and the percentage of live cells present were predicted. To determine a suitable method to perform the predictions, four analysis methods were investigated: SYTO 9:PI intensity ratio, adjusted dye ratio, single-spectrum SVR and multi-spectra SVR. We examined the applicability and performance of these analysis methods and propose a general optrode protocol for measuring percentage of live bacteria using multi-spectra SVR.

### Fluorescence profiles of SYTO 9 and PI

Both SYTO 9 and PI were used to stain the bacterial samples in this study, however, the fluorescence profiles of samples stained using only SYTO 9 or PI were also examined (Figure 1). The significantly higher SYTO 9 emissions in the presence of dead compared to live *E. coli* demonstrate the necessity of a counter-dye such as PI to enable quantitative fluorescence analysis of *E. coli* cells. Counterstaining may be required for most Gram-negative bacteria as the effect of stronger SYTO 9 staining of dead Gram-negative cells has also been reported in *Pseudomonas aeruginosa* [14]. It has been suggested that the strong SYTO 9 emission in dead cells could be due to its difficulty in crossing the outer membranes of Gram-negative bacteria [23], or that it is actively exported out of the cell [14]. Another possible explanation is that as these cells die, their nucleic acids become more accessible and loose as opposed to supercoiled, for SYTO 9 to bind to [24].

In the presence of PI, the SYTO 9 intensity in the dead *E. coli* samples decreased while the PI signal increased, compared to staining with the dyes individually, thus enabling the distinction between live and dead *E. coli* and subsequent quantification. Between the samples containing different composition of live and dead bacteria, there are some noticeable changes in the shape of the SYTO 9 peak emission. The SYTO 9 emission from the counter-stained 100% and 50% live samples show a subtle dual-peak at 498/501 nm, corresponding to the emission of the dye bound to DNA/RNA [25]. However, the dual-peak profile of SYTO 9 is not noticeable in the 100% dead sample. By accurately measuring the emission spectrum of counter-stained bacterial samples with the optrode, it is possible to obtain information about its composition from the dye intensities as well as the subtler changes in spectral shape. Initially, the dye ratio and adjusted dye ratio analysis were used to test the performance of simple methods that are primarily based on using the information about the peak dye intensities. Subsequently, to consider the information in peak dye intensities and the small changes in spectral shape, SVR was applied. SVR is a multivariate method that use information from the entire spectral window and thus can better characterise the spectral changes in relation to changes in the composition of the samples.

The photobleaching half-lives of SYTO 9 and PI were both found to be > 1s in their mB or mU states (Table 1). Thus, the effects of photobleaching on the spectrum measured in the first 20 ms is negligible. In addition, our results showed that the photobleaching half-lives of SYTO 9 compared to PI, and in the bound versus unbound states were not significantly different. While the photobleaching half-lives themselves may not provide sufficient information for distinguishing between SYTO 9 and PI or their binding states, they nonetheless represent useful time points for investigating spectral shape changes. As the bacterial composition of samples change, the ratio of bound to unbound SYTO 9 and PI also change, and this will influence the overall photobleaching rate of the samples. Thus, the extent of how much the signal has photobleached can provide additional clues about the proportions of live or dead bacteria present in the samples.

### Dye ratio and adjusted dye ratio

The dye ratio analysis is based on the method proposed in the product information guide of the LIVE/DEAD BacLight Bacterial Viability Kits [13]. However, as shown in Figure 2a, the relationship between the dye ratio and percentage of live bacteria only appears linear in a narrow range where samples contained low percentage of live cells. This is expected, considering the ideal behaviour of the dyes where SYTO 9 and PI stains live and dead cells, respectively. The percentage of live bacteria is equivalent to the number of live cells divided by the total cell number. In the dye ratio (Equation 1), the SYTO 9 intensity in the numerator is related to the abundance of live cells however, the PI intensity in the denominator does not represent the total cell number as PI should only stain dead cells. Following this observation, the adjusted dye ratio was derived (Equation 3), and its results are shown in Figure 2b. Although the relationship remains somewhat linear between the adjusted dye ratio and the percentage of live bacteria, the variations in results increase above *c.* 60% live. One reason for why the adjusted dye ratio is not linear across the entire range is due to a breakdown of the assumption that there is no interaction between the two dyes, e.g. fluorescence from SYTO 9 can excite PI [2]. Nonetheless, the adjusted dye ratio may be useful as a simple method for providing a rough indication of the proportion of live bacteria present in $10^8$ bacteria/mL bacterial samples.

### Single-spectrum and multi-spectra SVR

Multi-spectra SVR performed slightly better in modelling the training samples and evaluating the test set samples, as they returned lower standard error values than the single-spectrum SVR. Compared to single-spectrum SVR, more time information is included in the input to the multi-spectra SVR models which may be valuable to improving the predictive power of the SVR algorithm. The additional spectra used in multi-spectra SVR were chosen from measurements taken at times corresponding to the photobleaching half-life of the dyes in their bound or unbound states (Table 1), which provide time points for inspecting changes in spectral shape.

To find the spectral sequence for multi-spectra SVR that gives the lowest RMSE, several combinations of 2, 3 or 5 spectra obtained at different time points in the optrode measurement were concatenated and used as input to the SVR algorithm. The spectrum measured in the first 20 ms was included in all multi-spectra SVR analyses, as the difference in its signal intensity is mostly due to dye/bacteria interactions with minimal effects of photobleaching. In a few circumstances, the use of 2-spectra as the input obtained results with RMSE smaller than or close to that of the single-spectrum SVR. However, the use of 3 or 5 concatenated spectra obtained worse results than single-spectrum SVR (Table 2). An explanation for this is that the spectra obtained at subsequent additional time points do not add further compositional information and rather just introduce added spectral noise. For the multi-spectra SVR to be valuable, it requires a choice of time points for spectral input that is a balance between introducing new useful information and not adding redundant data.

The chosen input for multi-spectra analysis of $10^8$ bacteria/mL samples was a combination of the first spectrum and the spectrum recorded at the photobleaching half-live of mB-SYTO 9 (1244 ms). In these samples, the SYTO 9 peak intensity varies greatly according to the proportion of live bacteria present, and the spectrum taken at 1244 ms provides additional time information of this signal. On the other hand, the first spectrum and the spectrum recorded at the photobleaching half-life of mB-PI (1124 ms) was chosen for the multi-spectra analysis of $10^7$ bacteria/mL samples. There was little obvious change in the SYTO 9 or PI peak intensity in the $10^7$ bacteria/mL samples as its proportion of live:dead cells changed. However, in the $10^7$ bacteria/mL samples, the intensity of the PI peak was observed to dominate that of the SYTO 9 peak, by up to *c.* 7 times. Unsurprisingly, using the photobleaching half-lives of mB-SYTO 9 or mB-PI returned the lowest error in multi-spectra SVR, as the bound dye signals are a result of bacteria-dye interactions.

### Limitations and future improvements

As shown by the model validation results, the SVR models performed better in predicting the percentage of live bacteria in $10^8$ bacteria/mL samples compared to the $10^7$ bacteria/mL samples. The predictions for $10^8$ bacteria/mL samples struggle when there is a low proportion of live bacteria present, as shown by the test set sample containing 6.9% of live, which was not predicted within 2 standard errors of the 1:1 line. The model of $10^7$ bacteria/mL samples perform poorly and returns invalid predictions when the percentage of live bacteria is outside the measurable range of *c*. 7% to 73%. The SYTO 9 signal dominates the fluorescence emission when there is a high percentage of live bacteria, whereas the PI signal dominates when there is a high percentage of dead bacteria. Thus, making it more difficult to accurately obtain information from the dye signals at the extreme ends where there is either < 7% or > 73% of live bacteria present.

As the amount of dye used in this series of experiments was unchanged, the $10^7$ bacteria/mL samples contained significantly more unbound dye than the $10^8$ bacteria/mL samples. Thus, the signal to noise ratio is smaller in samples with lower bacterial concentration, due to the presence of a greater proportion of unbound dye. In an attempt to increase the signal to noise ratio, multiple spectra from one optrode measurement recorded from $10^7$ bacteria/mL samples were summed then used in analysis, however this did not improve the prediction results. One way to experimentally overcome this limitation is to decrease the concentration of dyes used in samples that contain low bacterial concentrations. This requires prior knowledge of the approximate bacterial concentration of the sample, and further investigations would be necessary to find the optimal dye volumes for different concentrations. Another option is to remove unbound dye from the samples via washing, however this requires extra sample processing and is difficult to complete without losing bacteria in the process [17]. Work is underway in our group to automate the dye staining and washing processes by using microfluidic platforms.

To further improve the accuracy for predicting the proportion of live bacteria present, the multi-spectra SVR analysis can be combined with principal components analysis (PCA). Already, we observed that the PCA scores of fluorescence spectra collected from blind samples can be used to inform which concentration group the samples belong to and accordingly, which SVR model to use for analysis. As shown in Figure 4, the spectra obtained from $10^8$ and $10^7$ bacteria/mL samples can be easily separated using the first two components obtained from the PCA. Recent work also demonstrated the possibility to predict the concentration of live and dead bacteria in a bacterial mixture, using principal components regression. By combining the information of bacterial concentration with the percentage of live bacteria present, it would be possible to obtain a more accurate and precise measurement of bacterial content in a sample.

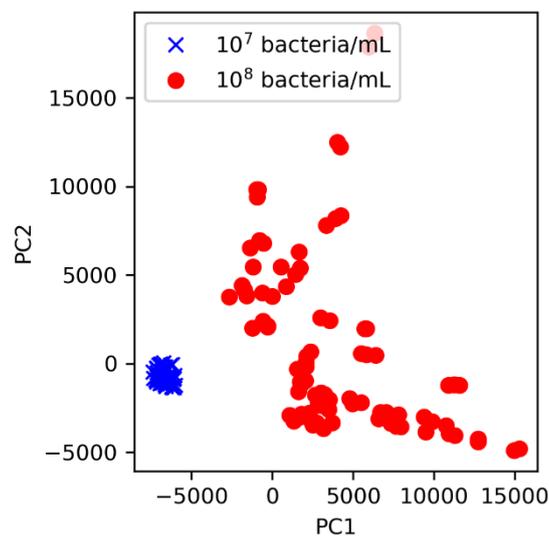

**Fig. 4** Principal component analysis scores plot of the first spectrum obtained from bacterial samples used to build the regression models. Although the samples contain varying live:dead bacterial cell ratios, there is a clear separation between samples that have a total concentration of $10^8$ and $10^7$ bacteria/mL.

## Conclusion

In this study we demonstrated the feasibility of using quantitative fluorescence spectra to monitor bacterial viability. We show this using spectral data obtained by the optrode, a convenient and cost-effective fibre-based spectroscopic device that accurately quantifies fluorescence signals in near real time. *E.* coli samples containing various proportions of live:dead cell concentrations were stained with SYTO 9 and PI, then subsequently measured using the optrode. Of the four analysis methods investigated, multi-spectra SVR obtained the most reliable results in predicting the percentage of live bacteria present. For $10^8$ bacteria/mL samples, the model was able to predict the percentage of live bacteria present down to *c.* 7% live. Predictions of the percentage of live bacteria in $10^7$ bacteria/mL was also achieved, though with a narrower detection range of *c.* >7% and <73% live. The main drawback from obtaining sensitive prediction at lower concentrations is the presence of unbound dyes, which can be reduced by decreasing the volume of dyes used or the use of an improved automated staining/washing process. This study demonstrates the value of an approach to improving the spectral processing method for monitoring the ratios of live:dead bacteria stained with SYTO 9 and PI. Further, we demonstrate that the multi-spectral SVR method and the optrode could be applied to monitor the effectiveness of a wide range of antimicrobial processes, including antibiotic treatment and food processing methods.

**Conflict of Interest:**

The authors declare that they have no conflict of interest.